# High-Contrast Chirped-Pulse Amplification Enabled by In-Band Noise Filtering

Jing Wang[1], Jingui Ma[1], Peng Yuan[1], Daolong Tang[1], Guoqiang Xie[1], Liejia Qian[1,*] and Frank W. Wise[2]

[1]Key Laboratory for Laser Plasmas (Ministry of Education), School of Physics and Astronomy, IFSA Collaborative Innovation Center, Shanghai Jiao Tong University, Shanghai 200240, China.
[2] School of Applied and Engineering Physics, Cornell University, Ithaca, New York 14853, USA
*email: qianlj19@sjtu.edu.cn

**Lasers that generate ultra-intense light pulses are under development for experiments in high-field and high-energy-density physics[1,2], as well as for applications such as particle acceleration[3]. Extensions to even higher powers are being considered for future investigations that can only be imagined today, such as the quantum electrodynamics of plasmas and isolated attosecond-pulse generation with solid targets[4,5]. For all of these areas, it is vital to produce high-contrast pulses, so that no pre-plasma is created in the target before the arrival of the main pulse[6,7]. However, noise is unavoidable in high-gain amplification, and is manifested in the form of background light that accompanies pulses generated by chirped-pulse amplification (CPA)[8,9]. Here, we introduce a linear filtering technique based on spatio-spectral coupling, which allows in-band filtering of amplified pulses for the first time. Experiments demonstrate approximately 40 times contrast enhancement in optical parametric chirped-pulse amplification (OPCPA)[9] and provide a foundation for scaling to much higher performance. The simplicity, efficiency, and direct compatibility with existing techniques for short-pulse generation will make spatio-spectral filtering attractive to a wide range of applications in ultrafast optics and time-resolved spectroscopy[10-12], and may open new directions in noise reduction.**

Noise is fundamental and ubiquitous in optics[13-16]. The evolution of noise that arises from spontaneous processes in optical amplifiers fundamentally limits the peak-to-background contrast of ultra-intense laser pulses. Extremely high contrast is required to avoid strong excitation of undesired effects by pre-pulses, pedestal, or background light that can accompany ultra-intense pulses[6,7]. Broadband noise can be efficiently filtered from narrowband signals in the frequency domain. However, signal and noise have similar spectral properties in amplifiers for ultrashort pulses, and the need for simultaneously efficient *and* high-contrast filtering compounds the challenge. Alternatively, noise can be suppressed in the temporal domain *via* nonlinear processes. Cross-polarized wave (XPW) generation can significantly attenuate amplified spontaneous emission (ASE), and improve the ASE-limited contrast of ultrashort pulses by several orders of magnitude[17,18]. The XPW technique removes in-band noise[15], and has enabled the achievement of contrast on the order of $10^9$ at hundreds of terawatts of power[19-21]. These attractive properties have led to wide use of the XPW technique for cleaning the millijoule-energy pulses that serve as the input or seed pulses for high-energy amplifiers. Multi-petawatt lasers need to achieve contrast levels[21] of $10^{10}$ to $10^{11}$. Nonlinear techniques such as XPW generation no longer suffice because the noise produced in high-gain amplification stages eventually dominates the pulse contrast, and such techniques are too lossy to be used after amplification. As a result, current petawatt (PW) lasers[22,23] typically achieve contrast of $10^8$ to $10^9$, which degrades high-field physics experiments such as the plasma acceleration of particles[3]. This situation is now becoming a major bottleneck to expansion of the frontiers of ultra-intense lasers and high-field physics.

Pulse cleaning at the amplification stage is therefore imperative[24]. This challenge calls for in-band spectral filtering, which can efficiently attenuate noise within the signal bandwidth. Unlike nonlinear techniques that are highly lossy and limited to low-energy unchirped pulses, linear in-band filters will not impair the signal and are compatible with high-energy chirped pulses. However, to the best of our knowledge, such a filter has not been demonstrated in optics.

Here we demonstrate the ability to linearly filter in-band noise in short-pulse amplification. Spatio-spectral coupling[25] underlies the filter concept, which we thus refer to as the spatio-spectral filter (SSF). The filter is simple, efficient, and naturally compatible with CPA as well as variants such as frequency domain optical parametric amplification[26]. As an initial demonstration of the SSF, we show that it can improve the contrast of high-gain OPCPA by a factor of 40 over the conventional OPCPA.

The SSF consists of a pair of parallel diffraction gratings with a slit at the midplane between the gratings (Fig. 1a). Let the first grating be illuminated by a polychromatic collimated light beam; the finite width of the slit ($W_s$) results in the selective transmission of narrowband components ($\lambda$) at corresponding diffraction angles ($\theta$). The transmitted light exhibits angular dispersion $d\theta/d\lambda$ and is subsequently converted into a spatially chirped ($dy/d\lambda$) beam by the second grating[27]. If diffraction is neglected, the transmission of the SSF can be written as

$$T(\lambda, y) = \begin{cases} 1, & for \ \left|y - \frac{dy}{d\lambda}(\lambda - \lambda_0)\right| \leq \frac{W_s}{2}cos\gamma_0 \\ 0, & otherwise \end{cases} \tag{1}$$

where $\lambda_0$ and $\gamma_0$ denote the central wavelength of the illuminating light and the angle of incidence with respect to the first grating, respectively. The spatio-spectral coupling coefficient is determined by the angular dispersion and the propagation distance ($L$) from the slit to the grating, $\frac{dy}{d\lambda} = L\frac{d\theta}{d\lambda}\frac{cos\gamma_0}{cos\theta_0}$. Owing to the coupling of the spectral and spatial coordinates, the slit produces a locally reduced spectral width (labelled in Fig. 1b) $\Delta\lambda_l = W_s cos\gamma_0 (dy/d\lambda)^{-1}$ in each spatial slice. In the same manner, the overall spectral transmission window ($\Delta\lambda_o$) is related to the incident beam size ($W_b$) by $\Delta\lambda_o = W_b(dy/d\lambda)^{-1}$. Thus, the overall window can be made sufficiently broad by use of a sufficiently wide incident beam. The SSF reduces to a conventional monochromator when the slit is placed at infinity: the coupling vanishes, and the overall window reverts to the locally-reduced spectral width. Measurements (Fig. 1c) confirm the calculated (Fig. 1b) transmission of an SSF with parameters appropriate to the noise-filtering experiments presented below. Broadband noise, which is randomly distributed in the spatio-spectral domain, will be mostly blocked. On the other hand, an incident broadband signal with spatio-spectral coupling matched to the transmission function will pass through the slit unimpeded. The reduction factor for in-band noise energy is approximately equal to the coefficient of the induced spatial chirp, $\Delta\lambda_o/\Delta\lambda_l$.

We applied the SSF to an OPCPA system. Parametric super-fluorescence (PSF) that can be comparable to the signal under saturation of the parametric conversion process leads to an incoherent noise pedestal in OPCPA[28,29]. The spatiotemporal chirp employed in OPCPA provides a natural way to implement the SSF: it can be incorporated in the pulse compressor (see Methods) by placing the second grating right behind the slit (Fig. 2a). A minor modification of the standard stretcher design yields the desired spatiotemporal chirp (see Methods). As a first step, we experimentally verified faithful spatiotemporal stretching and compression of ultrashort pulses without amplification. The incident Gaussian beam (Fig. 2b) was converted to a stripe-like spatio-spectral beam (Fig. 2c) in the process of stretching the seed pulse from 180 fs to 240 ps. The coupling coefficient of 2 mm/nm was matched to the SSF transmission shown in Fig. 1c. To ensure that the spatial chirp remained unchanged during beam propagation, an image-relay telescope was inserted between the stretcher and compressor. The compressor restored the stretched signal to its initial spatial (Fig. 2d) and temporal (Fig. 2e) profiles with high fidelity.

The unseeded parametric amplifier emits PSF with a typical bandwidth of more than 100 nm (Fig. 3a). Owing to the limited compressor passband, the PSF bandwidth was reduced to 64 nm. Closing the slit aperture activated the SSF. We spatially sliced the PSF beam in front of the compressor using a diaphragm to measure the spectra at the slit exit. The experimental results (Fig. 3b) clearly reveal the linear dependence of transmitted wavelength on position. In particular, the measured coupling coefficient of 2 mm/nm and the local bandwidth of 1.4 nm are consistent with their counterparts in the designed SSF (Fig. 1c) and stretcher (Fig. 2c). As a consequence, the PSF energy was reduced by a factor of approximately 40 (Fig. 3c), consistent with the ratio of compressor passband to the local spectral width of the SSF. The good agreement between the designed and measured noise reduction confirms the in-band filtering property of the SSF.

We assessed the performance of the SSF in controlled OPCPA experiments. Microjoule-energy seed pulses were amplified to the 2-mJ level, and compressed to full-width at half-maximum duration around 180 fs (see Methods for details). As shown in Fig. 4a, the amplified pulse contained a PSF pedestal within the pump-pulse window along with a flat noise background from the seed. The pulse contrast was therefore influenced by both types of noise, with the dependence on parametric gain shown in Fig. 4b. In the low-gain (<$10^5$) regime, the amplified pulses retained the $10^{11}$ contrast of the seed; this suggests that the seed noise dominated the pulse contrast. In the high-gain (>$10^5$) regime, the PSF noise overwhelms the seed noise. The performance of the conventional OPCPA set-up (see Methods) is severely compromised by PSF noise, with the contrast dropping well below $10^9$ at a gain around $10^6$. With the SSF, contrast above $10^{10}$ is achieved. The SSF produces about a factor of 40 improvement in contrast over the conventional OPCPA.

The generation of pulses with contrast that matches the seed contrast in the low-gain (<$10^5$) regime (Fig. 4b) is evidence of near-noiseless amplification in an OPCPA system. The SSF allows near-noiseless amplification to be achieved with gain an order of magnitude higher than the conventional OPCPA. The increased contrast obtained with high gain in this proof-of-concept demonstration of the SSF filter is already significant, but order-of-magnitude improvements should be possible with increased spatial chirp. Such performance would enable near-noiseless amplification in the high-gain (>$10^6$) regime, which implies the generation of high-contrast ($10^{11}$) 10-PW pulses from 100-GW level seed pulses. Therefore, the use of such an SSF with the planned 10-PW lasers[21] should facilitate clean laser-plasma interactions in the ultra-relativistic regime, with dramatic implications for particle accelerators and QED plasmas[3,4]. There is a continual push to shorter pulse durations, and the SSF will offer greater advantage with the corresponding increased bandwidths.

In combination with the CPA and XPW concepts, the SSF enables, for the first time, high-gain but low-noise amplification of ultrashort pulses. The straightforward integration of the SSF concept into existing designs, along with its efficiency, will aid its adoption. More broadly, the unique capability of in-band filtering should be valuable to a wide range of applications in ultrafast optics.

# Figures and captions

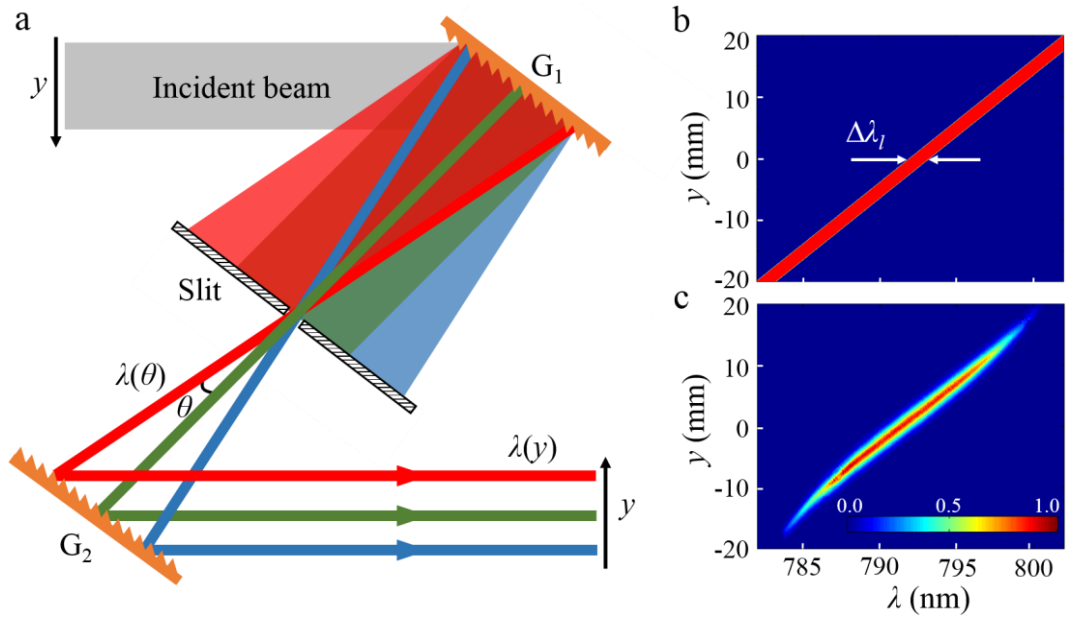

**Figure 1 | SSF and its transmission function. a,** Schematic diagram of the SSF, which consists of two diffraction gratings, $G_1$ and $G_2$. **b,** Transmission function calculated based on Eq. (1). **c,** Measured transmission function. For the calculation and experimental parameters, see Methods.

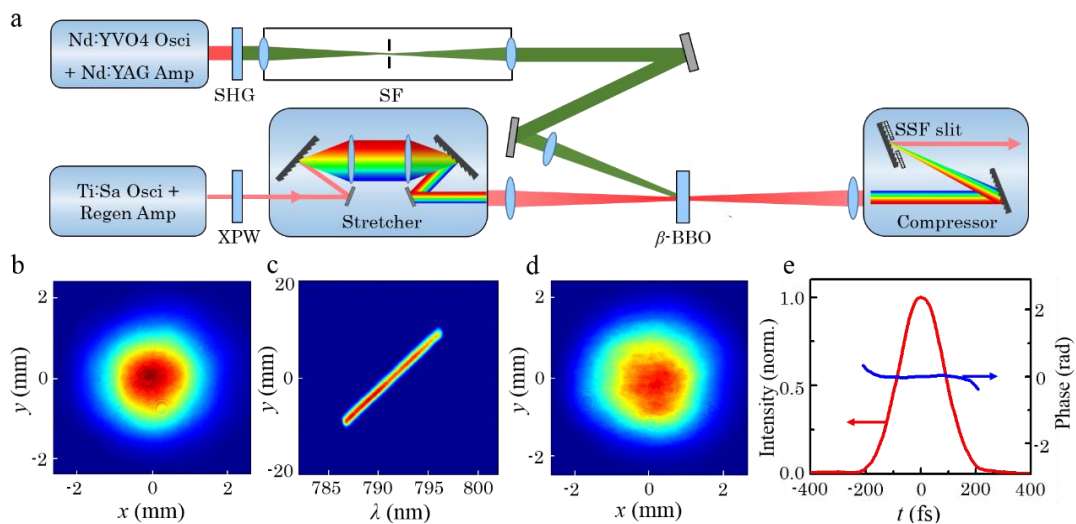

**Figure 2 | Characterization of a spatiotemporally chirped signal. a,** The OPCPA set-up using the SSF (see Methods for details). SHG, second-harmonic generation; SF, spatial filter; XPW, cross-polarized wave generation; $\beta$-BBO, beta-barium borate crystal. **b, c, d,** Recorded CCD images of the signal beam in the $x$-$y$ (before stretcher), $\lambda$-$y$ (after stretcher), and $x$-$y$ (after compressor)

domains, respectively. **e,** Temporal intensity (red) and phase (blue) of the compressed pulse measured via Frequency-Resolved Optical Gating (FROG).

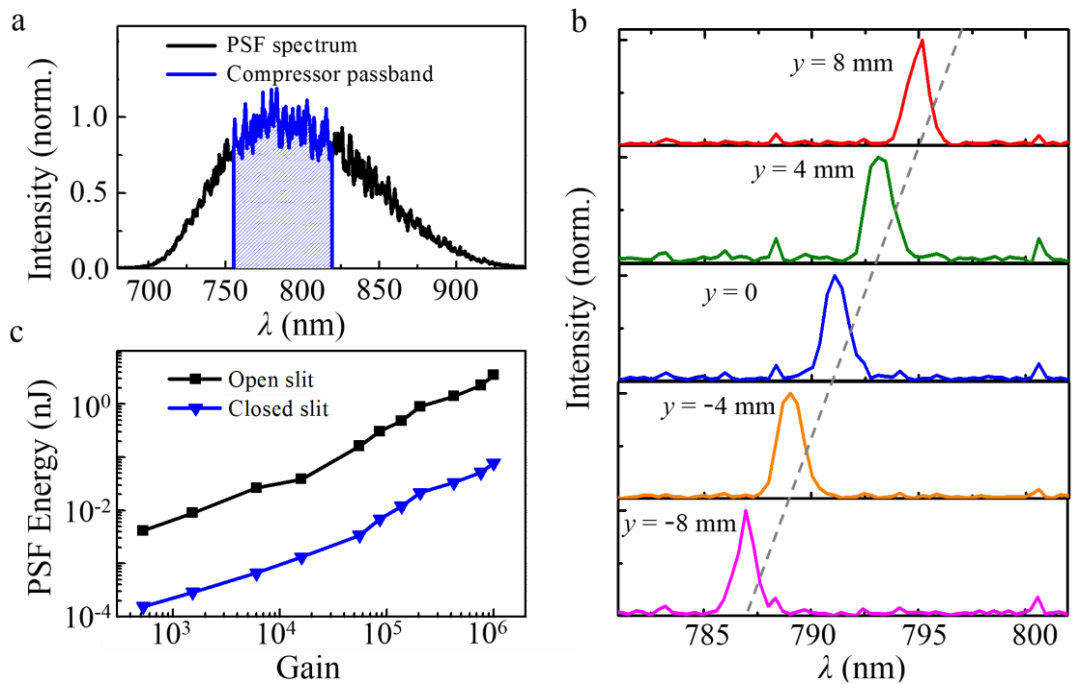

**Figure 3 | In-band filtering of the PSF noise. a,** PSF spectra measured before the compressor (black) and the compressor passband (blue). **b,** PSF spectra transmitted at the slit exit for five spatial slices of the incident beam on the compressor. These spatial slices were obtained by translating a 2.7-mm-diameter diaphragm along the y-axis. The dashed line, determined from the spectral-peak locations, represents the coefficient of spatio-spectral coupling. **c,** Measured PSF energy as a function of the parametric gain with the filter slit open (squares) and closed (triangles).

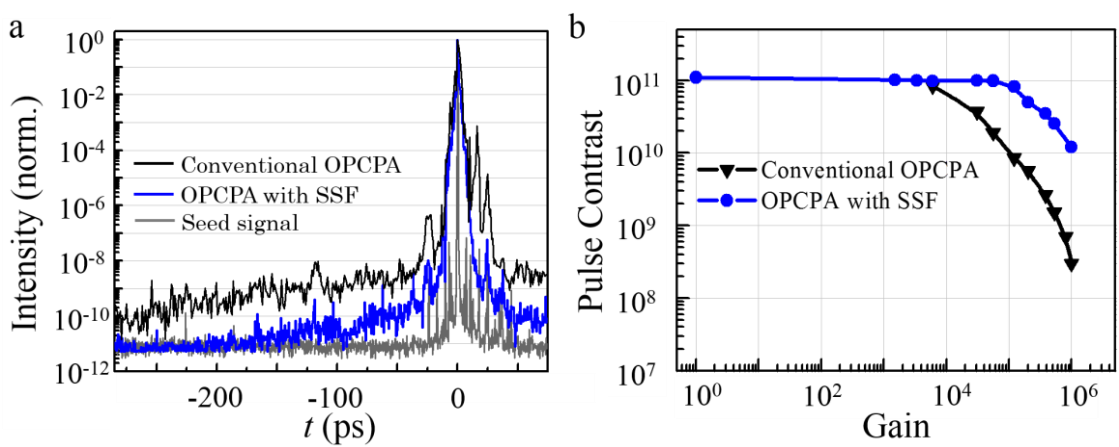

**Figure 4 | Pulse contrast measurements. a,** The cross-correlation traces of the amplified pulses at a parametric gain of $10^6$ in the OPCPA system with (blue) and without (black) the SSF. The seed pulse-contrast (grey) is $10^{11}$ for all times more than 10 ps from the peak. **b,** Pulse-contrast comparison under similar circumstances between the OPCPA system with the SSF (blue circles)

and a conventional OPCPA system (black triangles). Here, the pulse contrast is referred to the noise background at -50 ps.

## Methods

**SSF components and spatiotemporal compressor.** The SSF shown in Fig. 1a was constructed using two parallel gratings with a groove density of 1480 lines/mm and an aperture of 165 mm (Horiba Jobin Yvon, France). The filter slit, with a closed aperture of $W_s$ = 3.0 mm, was separated from both gratings by an equal slant distance of $L$ = 0.7 m. In the experiment, the filter transmission function was measured using the output from a commercial femtosecond Ti:sapphire regenerative amplifier (Legend Elite, Coherent, USA), which served as the necessary polychromatic light source. This probe light illuminated the SSF grating at an angle of incidence of $\gamma_0 =$ 26.1°. The specifications of the probe light are as follows: beam width, $W_b$ = 20 mm; central wavelength, $\lambda_0$ = 791 nm; and bandwidth, $\Delta\lambda$ = 30 nm.

The SSF depicted in Fig. 1a imparts both a spatial and a temporal chirp to the signal and thus can be used for pulse compression in OPCPA. To construct a compressor for the spatiotemporal signal, we placed the second grating and the filter slit together (Fig. 2a). The distance between the two gratings was thus changed to $L$ = 0.7 m, while the other parameters remained unchanged. In this configuration, the compressor has the same noise filtering functionality as does the SSF shown in Fig. 1a, but its output is an ordinary circular beam without any spatial chirp.

**Stretcher design.** To match the transmission function of the SSF, the stretcher should yield a spatiotemporal signal. Such a spatiotemporal stretcher can have a configuration similar to that of conventional pulse stretchers (Fig. 2a), but the signal passes through it only once. In this case, the output signal simultaneously acquires a spatial and a temporal chirp[25,27]. The incident signal in the $\lambda$-$y$ domain is assumed to be a Gaussian pulsed beam:

$$A_{in}(\lambda, y) = \exp\left[-2\frac{(\lambda-\lambda_0)^2}{(\Delta\lambda)^2}\right] \exp\left(-2\frac{y^2}{W_b^2}\right) \tag{2}$$

where $\Delta\lambda$ and $W_b$ denote the bandwidth and beam width, respectively. In deriving Eq. (2), we assume that the bandwidth is much less than the wavelength. The output from the spatiotemporal stretcher evolves into

$$A_{out}(\lambda, y) = \exp\left[-2\frac{(\lambda-\lambda_0)^2}{(\Delta\lambda)^2}\right] \exp\left[i\frac{kL}{2}\left(\frac{d\theta}{d\lambda}\right)^2 (\lambda-\lambda_0)^2\right] \exp\left\{-2\frac{\left[y-\frac{dy}{d\lambda}(\lambda-\lambda_0)\right]^2}{W_b^2}\right\} \quad (3)$$

The second term represents an induced group-velocity dispersion that results in chirped signal pulses. In contrast to conventional stretchers, spatial chirp arises, as indicated by the third term in Eq. (3). In particular, the spatial-chirp term is identical to the filter transmission function if the width of the incident beam is properly set. Consequently, the spatial chirp produced by the stretcher can be cancelled by the compressor once the temporal chirp has been compensated.

Note that the spatial chirp will be cancelled upon passing the signal back to the stretcher. Simultaneously, this two-pass stretcher doubles the temporal chirp. In our design, we adopt a three-pass configuration, such that the temporal chirp is tripled, while the spatial chirp remains the same as that for a single-pass stretcher. Accordingly, a three-pass spatiotemporal compressor configuration was also adopted in the experiment. In our stretcher design, the grating parameters and distance are the same as their counterparts in the compressor. In addition, the required telescope in the stretcher is composed of two lenses with identical focal lengths of 1 m. The spatial chirp coefficient for the experiments reported here was about 10. Stretchers and compressors with spatial chirp coefficient around 100 have been reported.

**OPCPA set-up.** The OPCPA pump laser consisted of, in sequence, a Nd:YVO$_4$ laser oscillator-regenerative amplifier system (Pico-Regen, High-Q, Austria), a 10-Hz Nd:YAG boost amplifier (SpitLight, Innolas, Germany) that generated pulses with Gaussian-like spatial and temporal profiles, and a 5-mm β-BBO SHG crystal. The 532-nm pump laser incident on the OPCPA crystal had a pulse energy of 15 mJ, a duration of 420 ps, and a beam diameter of 0.9 mm. The seed laser, originating from a femtosecond Ti:sapphire regenerative amplifier system (Legend Elite, Coherent, USA), was temporally cleaned using the XPW technique. An image-relay telescope with lenses of 1 m in focal length was inserted between the stretcher and compressor. After stretching, the 791-nm seed signal had the following characteristics: a pulse energy of 10 µJ, a bandwidth of 10 nm, a duration of 240 ps, and a beam diameter of 0.7 mm at the focus of the image-relay telescope. A type-I 10-mm-thick $\beta$-BBO crystal configured for non-collinear phase matching (with an intersection angle of 2.9° between the signal and the pump) was adopted in the OPCPA set-up. In the OPCPA experiments, the pump energies were adjusted from 5 to 13.5 mJ to obtain parametric gains that varied from $10^3$ to $10^6$, while the signal efficiencies were held fixed at approximately 15% by decreasing the signal energies from 10 to 0.1 µJ accordingly. The amplified chirped pulse with a flattop-shaped spectrum of 10 nm in bandwidth was restored to a nearly Fourier-transform-limited duration of ~180 fs after the compressor.

Two types of OPCPA systems were studied in this work: a system with an SSF and a conventional system. Except for the use of a conventional pulse stretcher and compressor in the conventional OPCPA system, the two systems were nearly identical (with, for example, the same chirped-pulse duration of 240 ps), and most components were shared between them, such as the pump laser, the femtosecond seed laser, and the OPCPA crystal.

**Pulse-contrast measurements.** All measurements were performed using a third-order cross correlator (Sequoia 800, Amplitude Technologies, France), which provides a high-dynamic range of $10^{11}$ at input energy >1 mJ.

## Acknowledgements

This work was partially supported by grants from the National Basic Research Program of China (973 Program) (2013CBA01505), the National Natural Science Foundation of China (11421064), the China Postdoctoral Science Foundation (2016M601577), and the US National Science Foundation (ECCS-1609129).

## Author contributions

L.Q. and J.W. conceived and discussed the original idea. L.Q. and F.W. supervised the project. J.W., J.M., D.T. and P.Y. designed and implemented the experiments. J.W. and L.Q. performed the calculations. L.Q., J.M. and G.X. supervised the experiments and participated in the data analysis. All authors contributed to the preparation of the manuscript.